# Bridging the Skills Gap: A Course Model for Modern Generative AI Education

**Anya Bardach, Hamilton Murrah**

Northwestern University

anyabardach@u.northwestern.edu, hamiltonmurrah@gmail.com

**Abstract**

Research on how the popularization of generative Artificial Intelligence (AI) tools impacts learning environments has led to hesitancy among educators to teach these tools in classrooms, creating two observed disconnects. Generative AI competency is increasingly valued in industry but not in higher education, and students are experimenting with generative AI without formal guidance. The authors argue students across fields must be taught to responsibly and expertly harness the potential of AI tools to ensure job market readiness and positive outcomes. Computer Science trajectories are particularly impacted, and while consistently top ranked U.S. Computer Science departments teach the mechanisms and frameworks underlying AI, few appear to offer courses on applications for existing generative AI tools. A course was developed at a private research university to teach undergraduate and graduate Computer Science students applications for generative AI tools in software development. Two mixed method surveys indicated students overwhelmingly found the course valuable and effective. Co-authored by the instructor and one of the graduate students, this paper explores the context, implementation, and impact of the course through data analysis and reflections from both perspectives. It additionally offers recommendations for replication in and beyond Computer Science departments. *This is the extended version of this paper to include technical appendices.*

## Introduction

In a short matter of time technology has fundamentally changed. With the release of ChatGPT in 2022, a wave of generative AI (GAI) tools have gained mass popularity, and the term "age of AI" is becoming increasingly commonplace in academic papers and Google search term trends.[1] Emerging research on the impact of GAI tools include Anthropic's report that university students predominantly use Claude for higher-order cognitive tasks and recent findings out of MIT that GAI usage can create "cognitive debt", replacing higher order cognitive function and diminishing performance on subsequent unassisted tasks (Anthropic 2025; Kosmyna et. al. 2025). This has led to hesitancy in higher education as to whether students should be permitted to use GAI tools. A review of guidelines put out by 50 universities found educators and administrators held positive sentiments toward GAI use in teaching practices but cautious regulatory sentiments toward student GAI use (An, Yu, and James 2025). With the dichotomy of personal adoption and academic restriction, we observe two increasing skills gaps for students…

**Industry versus higher education.** The 2024 Microsoft Work Trend Index highlighted that "66% of leaders say they wouldn't hire someone without AI skills," and further, 71% expressed preference for a less experienced candidate with AI skills than a more experienced one without (Microsoft 2024). There is promise for the newest generation of graduates if they are properly prepared to meet the current needs of AI and the workforce, with 77% of leaders saying new hires will be given greater responsibilities with the incorporation of AI (Microsoft 2024). Universities focus GAI restrictions on upholding academic integrity and mitigating cognitive impact. However, many also demonstrate a commitment to helping students achieve job-readiness by graduation. Reviewing mission and objective statements from top ranked U.S. universities revealed most included goals of "professional" or "career" success for students.

**Faculty versus students.** From a teaching lens, students are entering higher education with GAI habits already engrained. Because of this, AI restrictions in classrooms do not appear to prevent GAI use, instead risking students using GAI tools quietly with no formal guidance in increasingly punitive classroom environments. This predicament raises concern about degrading teacher-student relationships, which may bode for worse student outcomes than the tools themselves (Emslander et. al. 2025).

We argue that to maximize positive career and classroom outcomes, students need to know how to harness AI tools to their fullest capacities and they need universities to support and guide them in this. By sharing our findings from teaching a course on applications for existing GAI tools in software development, we offer one pathway for achieving this.

---

[1] https://trends.google.com/trends/explore?date=today%205-y&geo=US&q=age%20of%20AI&hl=en

## Course development

**Computer Science focus.** In this age of AI, few industries have seen as significant of an upside contrast as software engineering, the primary career path for Computer Science majors. 1.7% of all job postings require at least one AI skill in 2024 compared to 0.5% in 2010, but in computer and mathematical occupations AI skill requirement is 12.3% in 2024 compared to 1.6% in 2010 (Galeano, Hodge, and Ruder 2025). The advent and continual improvement of AI tools has provided developers with increased confidence and efficiency, and academics are calling higher education to prepare software engineering students for this reality (Ng et al 2024, Kirova et al 2024). Northwestern University's Computer Science department identified a need for classes where students are taught to use AI resources correctly and effectively, which produced the following objectives and structuring.

**Objectives.** The desired learning outcomes for the course were as follows:
- Give students a structured environment to test the capabilities of AI tools and understand their benefits and shortcomings
- Provide students with a course that would be applicable to the workforce
- Supply students with knowledge, talking points, and a completed project to highlight in employment interviews

**Curricular placement.** Part of the concern in permitting the use of GAI tools is students forgoing learning foundational knowledge in favor of completion speed. In the same way that we learn multiplication tables before we start using calculators, we believe it is vital that students understand what they are building even if a tool optimizes the process. Methods of integrating GAI tools into foundational courses while maintaining deep and meaningful learning will be the challenge of this generation of educators. Until a clear solution is found, we believe students should spend their first two years learning the foundations and building their programming skills before delving into GAI tools. To that end, the pilot implementation of this course was co-listed as a 3XX and 4XX level Computer Science course—available to graduate students and advanced undergraduates. The foundational 2XX course on data structures and algorithms was listed as a prerequisite. The goal was to target students who had deep prior knowledge of full-stack development and were in the process of transitioning to industry.

**Similar courses.** In creating the course, the teaching staff found only two other universities that had actively attempted to incorporate GAI in structural ways to their curriculum. At the University of Alabama at Birmingham (UAB), Amber Wagner, PhD, integrated GAI into a software design course and a senior capstone course (University of Alabama at Birmingham 2024). At Harvard University, lecturer Christopher Thorpe, PhD, created a new course rather than supplementing existing ones. It focused on the creation and evolution of a SaaS product in order to teach students about the phases of the software development lifecycle alongside GAI (The Harvard Crimson 2025). Our goal was for students to engage in coursework that they were familiar with and interested in, and that provided unique challenges, all while emphasizing the rapid building experience as opposed to structures and processes. As such our course was split into two phases. Inspired by Dr. Wagner's experience, we shaped the first two weeks as a crash course, assuming students come in with zero prior exposure to GAI software development. The remainder of the course focused on developing a full stack application in a capstone format akin to the Harvard model.

## Methods

**Structure.** This course was taught in ten weeks over 80-minute class sessions twice a week. The full topic schedule can be found in Appendix A of the extended version of this paper. 28 combined undergraduate and graduate students were enrolled in the course. It was offered through the Computer Science department as a special projects elective that could partially fulfill the project requirement for the undergraduate major. Grades were computed based on four components:
- Participation (10%). Class attendance was mandatory. Students were expected to participate in discussions on course topics and engage with guest lecturers.
- Industry research (20%). Students interviewed an individual in their area of interest and offered insights from the interview on the use of AI in their field.
- Assignments (50%). Five coding assignments: two foundational assignments and three iterative pair projects.
- Final Project (20%). The cumulative integration of the three pair projects, producing a full-stack application.

**Introductory Assignments.** As Dr. Wagner noted in the UAB Reporter (2024), developers engage in debugging, refactoring, and optimizing every day. Students worked with Python in two foundational assignments that together taught four key ways to use GAI in programming. This allowed them to quickly identify beneficial use cases, as well as establish a base level of confidence and comfort with the tools before jumping into larger bodies of work.
- *Assignment 1*. Test-driven development and code optimization. Students used Copilot or another GAI tool of their choosing for three tasks: 1. Write three functions using a provided test suite. 2. Write a test suite with over 90% coverage for a simple API provided in the source files. 3. Reduce runtime and optimize inefficient, unreadable code provided in the source files to be modular and readable.
- *Assignment 2*. Debugging and pseudocode. Students were provided with two buggy scripts. The first had easily

identifiable logic, computation, and syntax issues to be fixed with GAI. The second had more complex bugs that a GAI was less likely to immediately identify and was accompanied by a test suite. Students used GAI tools to debug the code and fix each failing test. Lastly, students experimented with supplying a GAI tool with different forms of pseudocode, from natural language descriptions to more formal pseudo syntax.

**Lectures.** Periodic, instructor-led lectures addressed topics relevant to the current assignment. This included best practices, development methodologies, common areas for optimizing development flow with GAI, and benchmarking data to help students select the best GAI tools. The instructor designed additional workplace simulations, holding "stand-ups," a customary practice in industry. He cold-called teams and asked for updates on what they had worked on, what they were planning to work on, any challenges they faced and how they solved them, any help they needed from their partner or the instructor, and their planned timeline. Students' decisions were regularly challenged to practice explaining their choice in approach. This forced them to deeply understand their application and why it was built the way it was and prepared them for situations they could face in the workplace. After each major assignment, the instructor reviewed submissions to find which teams differed in approaches and invited them to engage in friendly debate over the benefits and drawbacks of their choices. By forcing them to recognize and respond to direct challenges to their decision-making, students could gain more confidence in their own designs and decisions.

**Guest Lecturers.** Seven guest speakers offered a wide-ranging sense of expert perspectives on GAI use in different domains: a career senior managing director at a major software firm, a technical strategy consultant, an AI research and policy expert, a product manager involved with GitHub Copilot, a PhD student focused on the intersection of AI and ethics, a lawyer with an expertise in AI, and a software engineer in agentic AI research and development. The final speaker came twice, offering a live agentic coding demonstration the second time. These speakers offered a diverse set of opinions and information, and a window into how real GAI practitioners were using these tools in their lives and work. Most lecturers came to speak in person, with one joining virtually. Students asked a variety of questions and engaged in guided discussions and thought activities.

**Industry research.** Given the continued evolution of GAI tools and the diversity of uses across different companies and industries, part of a well-rounded education includes gaining a holistic view of how best practices can differ based on company size, revenue, industry, or even among individual employees. Students were tasked with reaching out to someone in the software industry to conduct an informational interview about GAI use at their company. The two main goals of this exercise were to give students first-hand experience talking with people in a field they are interested in and to have them collectively gather information on the multitude of ways that GAI is being used at different companies. Students submitted interview notes and participated in a class discussion on their insights.

**Final Project.** For the rest of the course, the students built a basic version of the popular social media platform X (formerly known as Twitter) in teams of two. Assigning a capstone project or cumulative task through which students can demonstrate their attainment and integration of the expected learning outcomes offers an opportunity for students to complete the course with a deep understanding of the ways the different skills they learned both are unique and interact. In pairs, they simulated a work environment with productive collaboration. Each pair chose their programming languages and frameworks, and were asked to justify these decisions in architecture documents (described below). The project was split into four assignments with no source code, and students were encouraged to push the capabilities of GAI. This allowed students to focus on the benefits and shortcomings of using GAI tools for each distinct component.

- *Backend development.* Sixteen functionality expectations, included aspects of account management, posting, user interaction, post interaction, and feed viewing. Students were allowed to combine functionalities into as many or as few functions as they liked. Students additionally built test suites with at least 80% coverage.
- *Frontend development.* Eight different expected page views with clear flow and functional navigation, including login, account, feed, and error pages. Students implemented access restrictions based on login status and inactivity timeouts as frontend security. Students built test suites with at least 60% coverage, a lower benchmark to account for the nuanced nature of testing frontend code.
- *Database development.* Using whichever system they felt fit best, support for seven aspects: profiles, tweets, likes, retweets, comments, blocks, and follows. Extra credit was offered for databases that supported posted images and profile pictures. Students created diagrams of their databases and populated them with mock data.
- *Full stack integration.* Students finished with end-to-end integration of the full stack, including backend access to the database and linking frontend actions to backend API calls. They built test suites with at least 50% coverage.

**Architecture Documents.** With each component, students wrote an architecture document, to encourage them to really think through how they were building their application. As AI tools continue to improve, it will be increasingly important for developers to have a strong understanding of architecture and be able to confidently make trade-off decisions. To that end, there was no structure given for how to write programs or what function names should be, and requirements were flexible with justifications. Emphasis was placed on why students chose what they did. There were no right or wrong answers, simply the expectation to explain every decision they made in an assignment, including their reasoning for their choices and against any alternatives they considered. This was the only portion of the class where they were explicitly asked not to use GAI. Formal bodies of writing were not required, provided there was coherent reasoning. This helped shift their focus from *what* they were building to *how* they were building it. The instructor emphasized that the knowledge moat that has surrounded software engineering for so long is evaporating with advancements in GAI. The expertise that cannot be replaced is the how and why behind code. Few software needs can be handled effectively without fundamental architectural understanding. As agentic coding improves and code generation becomes more accessible, we believe engineering will become exceedingly architecturally focused and prioritize individuals who can design optimal, tailored systems.

### Data Collection and Analysis

To assess the efficacy and value of the course, two mixed method surveys were conducted: a standard university course evaluation and a post-course survey. Time was allocated on the last day of instruction for completing the surveys. Their protocols are provided in Appendix B of the extended version of this paper.

**University course evaluations.** A general evaluation provided by the university to all students for each of their courses to gather feedback on course content and instructor performance. It collects demographic data, degree paths, class standing, and academic reasons for taking the course. Questions comprised of required standardized course metrics and optional open-ended prompts. The anonymized data were made available to instructors and other students who completed their respective course evaluations. This served as buy-in motivation for survey completion.

**Post-course survey.** In order to evaluate value and efficacy of the course, students were also required to fill out an anonymous quantitative measurement of agreement to seven statements on a scale from 1 (strongly disagree) to 7 (strongly agree), six multiple-choice questions (all with a fill-in other option), and thirteen open-ended questions.

## Results

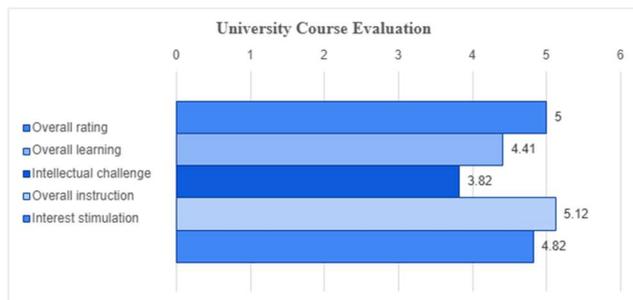

Figure 1: University Course Evaluation Averaged Scores.

**University course evaluations findings.** 17 of the 28 students submitted the university evaluation, yielding a 60.7% response rate. Eight were graduate students and nine were senior undergraduates, all Computer Science majors. One respondent reported not using the course to fulfill a degree requirement. Eleven students described their prior interest as a 6 (very high), and the remaining six a 5 (high). The estimated number of hours per week spent on coursework outside of class and lab time had a mean and median range of 4 to 7 hours, meeting university policy targeting two hours studying outside of class for every one hour in class.

Figure 1 displays the mean values for the five quantified questions on the survey. Students rated the overall quality of the course and the instruction highly, the overall quality of learning and the instructor's ability to simulate interest between 4 (somewhat high) and 5 (high), and the course's effectiveness in providing intellectual challenge between 3 (somewhat low) and 4 (somewhat high). Six affirmed that the course helped them learn, with one stating, "it was a unique, genuine learning experience," and another that "everyone at [the university] should take this." One student felt the course did not help them learn and said that the message they perceived was that "LLMs are only helpful at coding things when you already know how to do them," and "not having much experience with making a full stack application leaves you without much real help."

Thirteen students provided summaries on their experiences, which respondents are aware get released anonymously for other students to read. Twelve students gave positive reviews, three of whom included warnings to interested students to be clear that it is about learning "to use LLMs effectively while developing" and not meant for students new to full-stack development or looking for modern AI model implementations. One of these students noted "If you go into the class with the true understanding… you can take a lot out of it." Ten summaries, including the overall negative review, advertised the value of the guest speakers. One student said this was their favorite course they have taken at the university, and another declared in all caps "TAKE THIS CLASS." One student worried that the skills taught in the course were already out of date. One reviewer stated that the contents of the class were "not what [they] expected or wanted from the course," and that it was "difficult to succeed with such a fast-paced class without cutting corners with an LLM."

Five students offered suggestions for improving the course. One requested "more quantitative analysis of… different AI models." Another recommended letting students create any project as opposed to all building the same basic application. The third recommended larger groups to avoid their experience with an incommunicative partner, the fourth more thorough prerequisites to ensure students come in with development experience, and the fifth more guest lecturers from "Big Tech" companies.

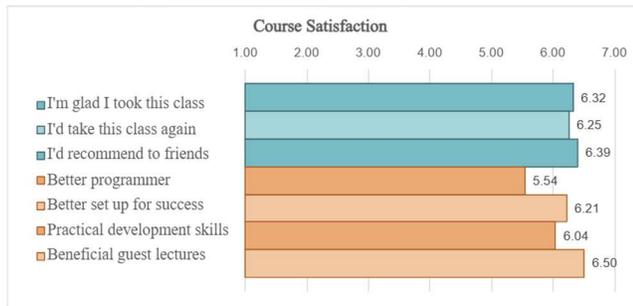

Figure 2: Post-Course Survey Averaged Scores.

**Post-course survey findings.** All 28 students took the tailored post-course survey. We divided the seven quantitative questions into two categories: course value and course efficacy. These are denoted in blue and orange respectively in Figure 2. No individual question had a mean or median score below 5 (Somewhat Agree), with most falling between 6 (Agree) and 7 (Strongly Agree). Whether students found the course valuable averaged at 6.32 on aggregate, and efficacy averaged at 6.07.

27 of the 28 students indicated the course increased their confidence in using GAI tools in real-world development, with the exception reporting "mixed" feelings. 14 students found GAI tools "very helpful", 13 found them "incredibly helpful", and one remained neutral. Several students reported doubts as to whether they would be able to apply GAI tools to their work in the companies they worked for or, in one case, their home country. Others expressed hesitancy over the value of "vibe coding." All 28 students reported having a confident individual approach to programming with AI.

Three students highlighted the industry research as the assignment that taught them the most and one student appreciated the second foundational assignment. The rest were evenly distributed across the four assignments building the full-stack application, with students preferring instances where they had to "really think through" and "get into the thick of it" when guiding the GAI process. One student said, "This class was the first that I not only had to really think about the choice I was making and why, but also be able to actually justify it in the report or during standup." Most students reported learning the least from the foundational assignments comparative to the others. Others felt the database assignment was the least involved and several recommended moving it prior to the backend assignment. 23 of 28 students preferred the split up nature of the final project into several assignments.

Regarding course structure, the most common comments were positive feedback on the guest speakers being "interesting," "cool," "beneficial," and "meaningful." Four students requested less speakers and three students requested more. There were additional requests to increase the number of hands-on demonstrations, as well as group discussions and debriefs on assignments. One student described how hearing their peers "un-stereotypical" approaches to their projects encouraged them to learn more by inspiring them to think outside the box. Students rated guest lectures highest of any aspect of the class, as noted in Figure 2.

The software engineer experimenting with agentic code who gave a live demonstration was the most popular, voted the favorite by 22 of the 28 students. Several students mentioned already incorporating his methods into their workflow, and one student said, "seeing someone working in action helped me think about what methods would work best for me." On the other hand, three students listed this speaker as their least favorite. All held concerns that his attitude toward agentic AI and vibing were "extremist" and one noted the speaker "raised some red flags about responsible use." The rest of the students were divided about their least favorite lecturer, predominantly citing their perception of the speaker's engagement level rather than the content of the lectures themselves. Students proposed a wide range of other lecturers they would have liked to see, including junior and senior developers in "Big Tech," coders in non-SWE fields such as biology, an environmental expert, and generally speakers with more hesitation or dislike toward GAI usage.

Of the 28 students, only 4 noted finding a portion of the class confusing, disorganized, or less valuable. Two felt the foundational materials were unnecessary, one felt they did not have enough foundational knowledge, and one requested more support in future iterations for writing test suites. There was one outlier across the dataset—a student who responded with values of 1 (Strongly Disagree) and 2 (Disagree) for all prompts save whether the course better set them up for success in the workforce, which they responded to with a 5 (Somewhat Agree). In their written responses they described how they "struggled from lack of prior experience" and had an uneven work dynamic with their project partner where they "didn't feel [they] could sufficiently contribute." All 28 students recommended the department should offer the course again, with one student using the fill-in other option to state, "Absolutely yes yes yes definitely yes."

## Discussion

**Student expectations.** One of the most noteworthy aspects of the data is the content of the student summaries. Unlike structured survey questions, these open-ended summaries represent the advice and warnings students shared explicitly for future cohorts to review. This intentional, peer-oriented discourse provides a unique lens into how students framed the value of the course. While the overwhelming majority of summaries were positive, many included explicit caveats underscoring that the course is not designed for students lacking prior development experience. Rather than detracting from the course's reputation, these warnings align with the pedagogical intent: the class was envisioned as a space to apply GAI tools within an existing foundation of software engineering skills, not as a shortcut to avoid learning those skills.

Other feedback across both surveys reinforces this point, as frustrations appeared to stem not from the GAI-centered content of the course itself but rather individual students' misunderstandings of course objectives, highlighting the importance of transparent communication about course expectations. Negative experiences stemming from mismatches between student assumptions and the actual scope of the course can be mitigated in future iterations by better clarifying in both the course title and the initial class session that students are expected to possess full-stack development experience and that the course is focused exclusively on applying existing GAI tools, not on learning modern model architectures. Additionally, clearer prerequisites will help ensure the course is positioned as intended: a capstone-like application education rather than an introductory programming or AI survey course.

**Application-based learning targets.** The relatively lower ratings on the dimension of "overall learning" actually appear consistent with the course's intentional design. The course aimed to be approachable and provide students with a scaffolded space to experiment with GAI tools, rather than presenting students with a steep new conceptual or technical learning curve. In this light, students' reports that the course was intellectually stimulating but not overwhelmingly challenging suggest an appropriate calibration of rigor. For a course whose central objective is tool fluency rather than deep technical model development, these outcomes are both expected and desirable. It is encouraging to observe student reflections such as the comment that "everyone at Northwestern should take this" that highlight the broader applicability of GAI literacy beyond the confines of Computer Science. Endorsements like this one support our argument that while this implementation of the course focused on full-stack software development, the underlying competency objectives—reasoning about and integrating GAI into workflows productively and responsibly—has relevance beyond software engineering. Understanding how to use GAI tools effectively is increasingly vital across fields.

**Guest speakers and ethics.** The consistently high value placed on guest speakers also merits discussion. Students not only enjoyed these external perspectives but also reported incorporating demonstrated practices into their own workflows, highlighting the previously unexplored value of authentic, practitioner-driven examples in GAI education.

At the same time, student feedback on diversifying the kinds of speakers to include non software developers, critical voices, experts from applied fields, and more suggests an opportunity to refine the integration of these lectures into the course structure to better balance variety and focus.

Students' concern about the agentic AI lecturer's extreme standards juxtaposed with his ranking as the favorite lecturer by a significant margin highlights a tension in AI education. Widespread use of AI appears to be here to stay but encouraging it may be hard to ethically justify in an increasingly unsustainable energy consumption climate. The methods and messages the agentic AI lecturer shared were highly impactful for students but gave little consideration to ethical implications at hand. Energy consumption levels of prompting an LLM are around 10 times higher than searching the internet, training an AI model can have the carbon footprint of five cars over their lifetime, and data centers can consume as much electricity as tens of thousands of homes (de Vries 2023; Hao 2019; Monserrate 2022). This focus on speed and innovation over impact threatens the objective of guiding students to responsible usage of AI tools. It will be vital for future iterations to include deeper discussion on the environmental impact of GAI usage. This was regrettable oversight in the first iteration.

Taken together, these key findings reinforce the viability and value of applied GAI education at the university level. Student enthusiasm was high, confidence in using GAI tools increased significantly, and critical feedback centered around refinements to strengthen alignment between course design and student expectations. The overwhelmingly positive reception, coupled with student calls for expanding access across disciplines, underscores the broader imperative of equipping the next generation of graduates across majors with GAI tool literacy. Our intervention successfully filled a critical gap: by integrating generative AI tools within a structured classroom setting, we transformed students' engagement from ad hoc experimentation to deliberate, ethical, job-aligned usage, and addressed the growing disconnect between industry expectations for AI fluency and reluctance in academic instruction.

## Experience Reports

**Educator perspective.** Looking back on the class, I consider it an enormous success. We took a nebulous idea and created what was by the vast majority of accounts a sincerely useful and informative experience. One reason it worked so well—not captured by the data—is because of the students who chose to enroll. They consistently came to class excited to learn and bought into class discussions, which made my role exceedingly easy. I was upfront with the class on the first day of what this course would and would not include. One student summary affirmed the value of this effort, noting that the course was "slightly different than [they] anticipated" but that "the professor makes it clear from day one... what the class is actually about." However, it is clear from the findings that this alone was insufficient for catching all unprepared students, and, in future iterations, I would modify the course prerequisites and description to be more explicit.

Delving further into future iterations based on the findings and teaching experience, I would let the students choose their own assignment builds. I would also incorporate more projects into the class, approaching something closer to rapid prototyping as opposed to one large project that is the focus of the entire course. The data makes it clear that the guest speakers set the course apart. I would bring in a few more of the different perspectives brought up by students to have a more diverse spectrum of thought.

Group creation is always going to be a possible pain point. At the beginning of the course, I put the choice of individual versus paired work to a vote, and the majority of the students present opted for pairs. Overall, it went very smoothly in our class, though some students' struggles with it are evident in the data. Whether individual projects or larger groups would improve the learning experience is worth discussing, and further data is needed to determine students' preferences.

**Student perspective.** I came to this course with extensive full-stack experience but little to no experimentation with GAI tools in my development process. Given the frequency and scope with which I had observed peers using these tools, I expected the course to be most useful in hearing about current industry practices and for the application aspect to be easy. Though not difficult, I was surprised to find that even the foundational assignments presented a bit of a challenge and required hands-on involvement. Indeed, throughout the course I found I had preconceived notions that using GAI was an all or nothing choice and would feel like giving up my power in the process. This sentiment was steadily corrected as the process helped me see that GAI tools are just that—tools. By the end, I felt empowered in my ability to apply GAI to my work while still feeling like it was my own.

I believe that this is a vital process for students from all disciplines to experience in the search for healthy and productive relationships with GAI. The strategy that I have found most useful in my personal workflow is using GAI as a sounding board for building airtight specification documents. The process was deceptively simple: prompt an LLM with a broad idea and have it ask you iterative, binary questions to flesh out a fully detailed specification. Engaging in this process has helped me think more deeply about my development plans, mitigating sunk cost fallacy and significant backtracking. In future iterations of this course, I would echo other suggestions to move the database assignment before the backend assignment to better align with a full-stack workflow. I think giving students the freedom to choose their own projects is a promising idea so long as space is still made for students to compare their different strategies and tool recommendations.

I took this course because I was both highly wary of incorporating generative AI tools into my workflow and highly aware of the fact that this was putting me at an academic and job market disadvantage. While I did not want to violate academic policy, the sheer quantity of people around me I saw disregarding restrictions made it increasingly clear that the policies were ineffective in the face of the attraction to the tools. I was excited about this course, as it provided an option for exploring these tools without fear of punitive measures. I found the course to be really useful in this regard and consistently engaging. I now feel confident using GAI tools to expand on other project ideas I have had and bolster my portfolio. I incorporated the practices we were learning into my thesis project and am using them to inform research on how GAI can empower educators to shift away from traditional pedagogies toward learner-centered practices.

## Conclusions

A novel course was implemented to address the importance of teaching students to effectively use generative AI tools for their academic wellbeing and job readiness. Through a mixed-method evaluation, we collected and analyzed rich data detailing students' experiences, perceptions, and preparedness in engaging with these tools. Participants reported significantly increased confidence in employing AI tools, deeper understanding of their strengths and limitations, and a clearer alignment with workforce expectations in software development contexts. These outcomes underscore the broader implications for higher education: rather than discouraging AI tool usage through restrictive policies, institutions should reframe classroom instruction to support informed, ethical engagement with emerging technologies. Embedding generative AI competencies into coursework not only bridges the disconnect between industry demand and academic policy but also reinforces universities' commitment to cultivate job-ready graduates.

**Future directions.** Computer Science felt like a natural starting point for experimenting with GAI instruction, but these tools are being used ubiquitously across fields. To build on this work, we recommend adapting the course model to different disciplines by substituting software engineering for other impacted career paths.

Though there were few comparable examples at the time of developing this, other institutions are simultaneously engaging in similar work. A quick search for GAI courses will generate results from Stanford, Johns Hopkins, Cornell, Carnegie Mellon, and more—mostly online mini-courses rather than integrated curricula. Future research should explore multi-institutional adoption, assess long-term impacts of different implementations on post-graduation outcomes through longitudinal study, and develop frameworks to ensure equitable access to GAI literacy—including open-source or institutionally supported access to cost-restrictive tools.

## Appendix A: Course Schedule

| Week | Topic | Description |
|---|---|---|
| 1 | Introduction | Provide an overview of the course and its main objectives |
| 2 | TDD + GenAI Foundations | Emphasize the value of test-driven development in a GenAI workflow; Discuss and practice basic concepts of using GenAI in software development, focusing on common use cases like bug fixes and test writing |
| 3 | Backend Development | Provide students with frameworks, best practices, and an understanding of common pitfalls to avoid when developing the backend of their full-stack application |
| 4 | LLM Landscape | Share a view of the top existing LLMs and their pros and cons as well as the best use cases for each |
| 5 | Company Stances on AI + Frontend Development | Student led discussion on the rules, regulations, and future direction of companies as it relates to AI and GenAI; Provide students with frameworks, best practices, and an understanding of common pitfalls to avoid when developing the frontend of their full-stack application |
| 6 | Vibe Coding | Provide students with a clear perspective and example of the extremes of GenAI usage in software engineering |
| 7 | AI Ethics | Raise potential concerns and topics to consider with GenAI largely and on an individual developer basis |
| 8 | Database Development + Real Applications of AI | Provide students with frameworks, best practices, and an understanding of common pitfalls to avoid when developing the database of their full-stack application; Showcase real examples of AI projects and capabilities that exist across different companies today |
| 9 | Legal Implications of AI | Showcase the potential for legal backlash and consequences for a software developer that utilizes GenAI in their workflows |
| 10 | Product Management | Provide an overview of the product management career trajectory, the types of work it involves, and the traits that make someone well-suited for it |

## Appendix B: Survey Protocols

**Post-Course Reflection Form**

**Quantitative section**
- I'm glad I took this class
- If I could go back, I'd take this class again
- I'd recommend this class to my friends if it were offered
- This class made me a better programmer
- This class set me up for more success when I enter the workforce compared to when I enrolled
- This course helped me develop practical software engineering skills
- Did you find the guest speaker sessions beneficial and enjoyable?

**Set-choice section**
- Did this course increase your confidence in using AI tools in real-world development?
    - Absolutely
    - Definitely not
- How helpful were AI tools in completing your assignments?
    - Incredibly helpful – I don't want to build without them ever again
    - Very helpful
    - Neutral
    - Not very helpful / I could've done the assignments faster on my own and don't plan to continue using many AI tools
- Are you leaving this class with a confident mental plan for how to effectively program with AI?
    - Absolutely
    - Not at all
    - If yes, students were asked to briefly describe the approach they walked away with
- Did you like splitting the project into smaller assignments? Or would you have preferred everything together?
    - Split up
    - Together
- How did the number of guest speaker sessions feel?

- I wanted more
- I wanted less
- The number we had felt right
- Should the department offer this course again?
  - Yes
  - No

**Qualitative Section**
- What are the top 2-3 things you learned in this course that you'll use in your career?
- What are the top 2-3 things that you think were not applicable to you?
- Which assignment taught you the most / was the most beneficial? Why?
- Which assignment taught you the least / was the least beneficial? Why?
- How did you feel about the class structure (Mix of lectures, discussions, debates, group breakouts)? Did this
- mixture facilitate learning? Inhibit it? Would you have changed anything?
- Is there anything the professor could have done differently to support your learning and make the entire class
- more valuable for you?
- Which guest speaker was your favorite? Why?
- Which guest speaker was your least favorite? Why?
- Are there any types of guests or industries you would've liked to hear from that we didn't?
- Were there any topics or tools you wish had been covered (or covered in more depth) that weren't?
- Did any parts of the course feel too fast or too slow?
- Were there moments that felt confusing, disorganized, less valuable, or just generally a waste of time?
- Please share any final thoughts you have on the class here

### University Course Evaluation

**Quantitative section**
- Subject interest before taking the course on a scale of 1 (not interested at all) to 6 (extremely interested)
- Average number of hours per week spent on coursework outside of class and lab time with options starting at "3 or fewer" and proceeding at 3 hour intervals until "20 or more".
- Overall course rating on a scale of 1 (very low) to 6 (very high)
- Self-assessed learning on a scale of 1 (very low) to 6 (very high)
- Efficacy in being intellectually challenging on a scale of 1 (very low) to 6 (very high)
- Overall rating of instruction on a scale of 1 (very low) to 6 (very high)
- Instructor's efficacy in stimulating interest on a scale of 1 (very low) to 6 (very high)

**Qualitative section**
- Did the course help you learn? Why or why not?
- Please summarize your reaction to this course focusing on the aspects that were most important to you.
- Efficacy in being intellectually challenging on a scale of 1 (very low) to 6 (very high)
- Overall rating of instruction on a scale of 1 (very low) to 6 (very high)
- Instructor's efficacy in stimulating interest
- What are the primary teaching strengths of the instructor?
- What are the primary weaknesses, if any, of the instruction?
- Can you offer suggestions for improvement?